\documentclass[aps,prb,groupedaddress,twocolumn]{revtex4-2}

\usepackage{booktabs}
\usepackage{siunitx}
\usepackage{amssymb}

\usepackage{graphicx}
\usepackage{amsmath}
\usepackage{amssymb}
\usepackage{hyperref}

\newcommand{\avg}[1]{\left<#1\right>} 
\renewcommand{\vec}[1]{\ensuremath{\mathbf{#1}}} 
\newcommand{\ket}[1]{\big| #1 \big>} 
\newcommand{\lr}[1]{\left(#1\right)}
\newcommand{\tot}[1]{#1_{\textsubscript{tot}}}

\newcommand{\cC}[1]{c^{\dagger}_{#1}}
\newcommand{\cA}[1]{c_{#1}}

\begin{document}

\title{Persistent Haldane Phase in Carbon Tetris Chains}
\author{Anas Abdelwahab}
\affiliation{Leibniz Universit\"{a}t Hannover, Institute of Theoretical Physics,  Appelstr.~2, 30167 Hannover, Germany}
\author{Christoph Karrasch}
\affiliation{Technische Universit\"{a}t Braunschweig, Institut f\"{u}r Mathematische Physik, Mendelssohnstraße 3, 38106 Braunschweig, Germany}
\author{Roman Rausch}
\affiliation{Technische Universit\"{a}t Braunschweig, Institut f\"{u}r Mathematische Physik, Mendelssohnstraße 3, 38106 Braunschweig, Germany}

\date{\today}

\begin{abstract}
We introduce the concept of ``tetris chains'', which are linear arrays of 4-site molecules that differ by their intermolecular hopping geometry. We investigate the fermionic symmetry-protected topological Haldane phase in these systems using Hubbard-type models. The topological phase diagrams can be understood via different competing limits and mechanisms: strong-coupling $U\gg t$, weak-coupling $U\ll t$, and the weak intermolecular hopping limit $t'\ll t$. Our particular focus is on two tetris chains that are of experimental relevance. First, we show that a ``Y-chain'' of coarse-grained nanographene molecules (triangulenes) is robustly in the Haldane phase in the whole $t'-U$ plane due to the cooperative nature of the three limits. Secondly, we study a near-homogeneous ``Y$^{\prime}$-chain'' that is closely related to the electronic model for poly(p-phenylene vinylene). In the latter case, the above mechanisms compete, but the Haldane phase manifests robustly and is stable when long-ranged Pariser-Parr-Popple interactions are added. The site-edged Hubbard ladder can also be viewed as a tetris chain, which gives a very general perspective on the emergence of its fermionic Haldane phase. Our numerical results are obtained by large-scale, SU(2)-symmetric tensor network calculations. We employ the density-matrix-renormalization group as well as the variational uniform matrix-product state (VUMPS) algorithms for finite and infinite systems, respectively. The numerics are supplemented by analytical calculations of the bandstructure winding number.
\end{abstract}

\maketitle

\section{\label{sec:Introduction}Introduction}

The ground state of the $S=1$ quantum spin chain is a prototype of a symmetry-protected topological (SPT) state and a representative of the gapped Haldane phase~\cite{Haldane1983,AKLT1988}.
In this state, each spin-1 breaks apart into two spins-$1/2$, which form singlets to the left and right and result in a valence-bond spin liquid that does not break the translational symmetry of the lattice. For open boundary conditions (OBC), two spins-$1/2$ at the ends remain as exponentially localized magnetic edge states.

Unlike topological insulators, where SPT states arise due to the nontrivial topology of the bandstructure, the spin chain is fundamentally an interacting system without a free-particle limit.
The Haldane phase is by now well-understood within the spin-only model~\cite{KennedyTasaki1992,Pollmann2010,Pollmann2012,Verresen2017}. However, an interesting question is how it can emerge in more general fermionic systems with charge fluctuations~\cite{Anfuso2007,Moudgalya2015}, with the possibility of additional unconventional states~\cite{Rausch2020,JuliaFarre2022} or phase transitions~\cite{He2024}.

There are several known mechanisms to connect interacting fermionic quasi-1D models to their counterpart $S = 1$ spin models, which we will now recapitulate.

\paragraph*{Mechanism 1:} The most direct way is a multiorbital chain, where two electrons strongly interact ferromagnetically via atomic Hund's rule and form $S=1$ objects, with the low-energy sector of the lattice being given by the spin model. A typical example are Ni-based compounds~\cite{Wierschem2014} with two electron holes in the 3d shell for every Ni$^{2+}$ ion. Interestingly, the Haldane phase remains robust even when the local magnetic moments are not fully developed~\cite{Jazdzewska2023}. We do not consider this route in this paper.

\paragraph*{Mechanism 2:} Hund's rule is also valid for most molecules~\cite{Katriel1977}: If single-orbital atoms are coupled to a molecule such that the highest occupied molecular orbital (HOMO) is degenerate with two electrons in it, a triplet $\tot{S}=1$ ground state is selected in most cases if interactions are introduced. For example, the level structure of an equilateral triangle is given by a nondegenerate level followed by a twofold degenerate one. When filled by four electrons (2/3 filling) and endowed with a weak interaction (such as a local Hubbard interaction $U$), the two electrons in the HOMO align ferromagnetically~\cite{Nourse2021}, and a chain of weakly coupled molecules exhibits the Haldane phase, a situation suggested for Mo$_3$S$_7$(dmit)$_3$~\cite{Janani2014,Nourse2016}. As for mechanism 1, the Haldane phase remains robust even if the total spin is not well-defined. We note that another material class with $\tot{S}=1$ molecules are porphyrin chains~\cite{Jelinek2023,Hamamoto2023,Zhao2023}.

\paragraph*{Mechanism 3:} A similar route is provided by Lieb's theorem~\cite{Lieb1989}, which states that on a bipartite Hubbard cluster with sublattices A and B that fulfill sublattice symmetry (i.e., hoppings exist only between the sublattices, not within), the ground-state spin at half-filling is given by the sublattice imbalance
\begin{equation}
\tot{S} = \frac{1}{2}\big|N_A-N_B\big|.
\label{eq:LiebStot}
\end{equation}
To understand this intuitively, one can first go to the strong-coupling limit where one effectively has a Heisenberg two-site spin model of the form $\vec{S}_\text{A}\cdot\vec{S}_\text{B}$ with $\vec{S}_{\text{A(B)}}=\sum_{i\in\text{A(B)}} \vec{S}_i$, so that the total spins of the respective sublattices are coupled. When the Hubbard interaction is reduced, no phase transition takes place and Eq.~\eqref{eq:LiebStot} in fact holds for arbitrarily weak interactions.
An important example of this mechanism are the triangulene nanoflakes where the sublattice imbalance results from an appropriate cutout from graphene~\cite{Morita2011,Su2019,Su2020,Turco2023}. When coupled as a chain, they provide an especially clean implementation of the Haldane phase due to the absence of parasitical interchain couplings that are present in bulk materials~\cite{Mishra2021,Martinez-Carracedo2023,Catarina2022,Henriques2023}.

\paragraph*{Mechanism 4:} A fourth route is provided by taking a chain with a topologically nontrivial band structure such as the (spinful) Su-Schrieffer-Heeger (SSH) chain and endowing it with a weak Hubbard interaction~\cite{Manmana2012,Yoshida2014,Sbierski2018,Rachel2018}. The emergence of the Haldane phase can be motivated by looking at the edge states for open boundary conditions. The spin-degenerate edge states will split upon an arbitrarily weak Hubbard coupling $U$; the low-energy singly occupied states remain degenerate and coincide with the spinful edge states of the Haldane phase, while the empty and doubly occupied states receive an energy penalty from the Hubbard term. In the bulk, the SSH dimers evolve into the singlet valence bonds of the Haldane phase. For $U\to\infty$, the system evolves into a dimerized spin-$1/2$ chain, which also features a Haldane phase \cite{PhysRevB.45.2207}. The emergence of the Haldane phase in the interacting SSH chain can therefore be understood from both the weak- and strong-coupling limit~\cite{mik22}.

\paragraph*{Mechanism 5:}\label{par:Mechanism5} A more subtle case is the Hubbard ladder where the noninteracting bandstructure is gapless but which for finite $U$ undergoes a topological phase transition to a nontrivial phase if the inter-chain hopping is increased~\cite{Nonne2010,Verresen2017,Sompet2022}. Starting in the gapless phase, an arbitrarily small Hubbard interaction induces a gapped Mott insulator and prefers singlet formation along the rungs that correspond to the valence bonds of the Haldane phase. If two edge sites are attached, they do not form singlets and instead develop into magnetic edge states. Such a realization of the Haldane phase has been observed in an optical lattice~\cite{Sompet2022}.

\paragraph*{Mechanism 6:} In the strong-coupling Heisenberg limit of large interactions, we generally obtain spin-$1/2$ systems that can, however, exhibit the Haldane phase if the underlying geometry is more complex than a simple chain. A prime example is the diagonally-coupled $S=1/2$ Heisenberg ladder, whose ground state is smoothly connected to the ground state of the $S=1$ spin chain~\cite{White1996}.

\begin{figure*}
\begin{center}
\includegraphics[width=0.75\textwidth]{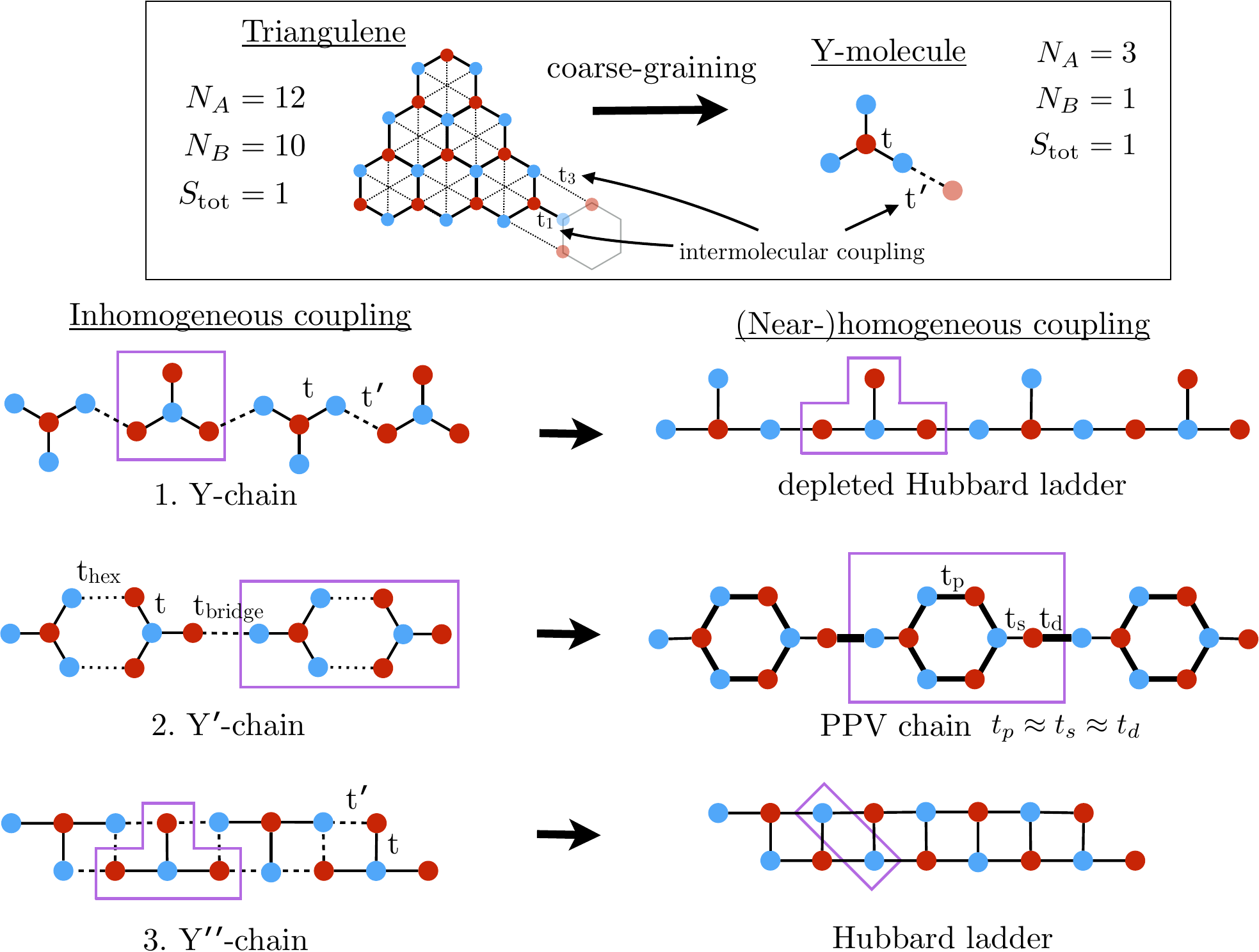}
\caption{
\label{fig:TetrisSketch}
Sketch of the geometries considered in this paper. The triangulene molecule with $N=22$ sites and hoppings $t_1$, $t_3$ can be coarse-grained into an effective Y-shaped molecule with $N=4$ sites, hoppings $t$, and an effective intermolecular hopping $t'$. The blue (red) dots denote the majority (minority) sublattice sites that are otherwise equivalent.
Coupling these Y-molecules produces different ``tetris chains'':
1.) A Y-chain which approximates a chain of triangulene and becomes a depleted ladder in the homogeneous limit $t\approx t'$.
2.) A Y$^{\prime}$-chain which in the near-homogeneous limit is related to a chain of PPV polymers.
3.) A Y$^{\prime\prime}$-chain which becomes the Hubbard ladder in the homogeneous limit.
In all cases, the purple box indicates the physical unit cell, and the figure also illustrates how the chains are being terminated for open boundary conditions.
}
\end{center}
\end{figure*}

In this paper, we first study a coarse-grained model of triangulene (the ``Y-chain'', see Fig.~\ref{fig:TetrisSketch}), which has been previously investigated in detail in the limit of weak intermolecular couplings $t' \ll t$, leading to the Haldane phase via mechanism 3~\cite{Catarina2022,Henriques2023}. We show that the Haldane phase is in fact stable for arbitrary values of $t'$ and $U$. This is due to the fact that (a) the bandstructure is topological, and (b) the strong-coupling Heisenberg limit features the Haldane phase for any value of $t'$. Thus, mechanisms 3, 4, and 6 conspire in this case.

We then explore a different geometry (the ``Y$^\prime$-chain'', see Fig.~\ref{fig:TetrisSketch}) that in the near-homogeneous hopping limit is related to the poly(p-phenylene vinylene) (PPV) polymer~\cite{Burroughes1990,Soos1993,Chandross1997A,Chandross1997B,Chandross1997C,Chandross1997D,Bursill1998,Bursill2002,Muellen2023}. We find that the mechanisms 4 and 6 now compete with each other, and the Haldane phase emerges only in certain parameter regions. However, we find that around the PPV point, the system is in fact robustly in the Haldane phase for any value of the Hubbard interaction $U$. We thus predict that PPV polymers should exhibit $S=1/2$ edge states if two carbon atoms are attached to the edges. To the best of our knowledge, this has not been noticed up to now despite intensive past research on the electronic model for PPVs. We further demonstrate that the Haldane phase is stable towards longer-ranged Coulomb interactions within the Pariser-Parr-Popple (PPP) model.

The coarse-grained triangulene model, PPV polymers, and the site-edged Hubbard ladder can all be subsumed under the category of \textit{tetris chains} and only differ by their intermolecular couplings (see Fig.~\ref{fig:TetrisSketch}). We believe that this is a useful framework to systematically find or engineer the Haldane state in real materials.

\section{\label{sec:Model}Models}

\subsection{Hubbard model}

The systems studied in this paper are goverened by a general tight-binding Hamiltonian
\begin{equation}
H_0 = -\sum_{i<j,\sigma} t_{ij} \lr{c^{\dagger}_{i\sigma}c_{j\sigma}+\text{h.c.}}.
\label{eq:H0}
\end{equation}
The operator $c^{\dagger}_{i\sigma}$ creates an electron with spin projection $\sigma=\uparrow,\downarrow$ at site $i$. The parameters $t_{ij}$ encode the hopping structure between nearest neighbors, which is best conveyed graphically, see Fig.~\ref{fig:TetrisSketch}. The corresponding unit cells of length $L_{\text{cell}}$ are shown as purple boxes in Fig.~\ref{fig:TetrisSketch}. We denote $N_{\text{cells}}$ and $L=L_{\text{cell}}\cdot N_{\text{cells}}$ as the total number of unit cells and the total number of sites, respectively.

A local Hubbard interaction of strength $U$ is modelled by
\begin{equation}
H_{\text{Hub}} =H_0+ U \sum_{i} \lr{n_{i\uparrow}-\frac{1}{2}}\lr{n_{i\downarrow}-\frac{1}{2}},
\label{eq:HHub}
\end{equation}
where $n_{i\sigma}=c^{\dagger}_{i\sigma}c_{i\sigma}$ is the particle number density. The total particle number is given by
\begin{equation}
N_{\text{tot}}=\sum_{i} \avg{n_{i}},~~~n_i=\sum_{\sigma}n_{i\sigma}.
\end{equation}
We also introduce the spin operator as well as the total magnetization
\begin{equation}
S^z_i=\frac{n_{i\uparrow}-n_{i\downarrow}}{2},~~M_{\text{tot}} = \sum_i \avg{S^z_i}.
\end{equation}
For carbon, the typical energy scales are $t\sim2-3\,\text{eV}$, $U\sim8-11\,\text{eV}$~\cite{Wehling2011,Barford2013}, so that the ratio $U/t\sim2-5$ puts such systems into the weak-to-intermediate coupling regime.

The Hubbard model features a spin-SU(2) symmetry with quantum numbers $S_\text{tot}$ and $M_\text{tot}$ as well as a charge-SU(2) symmetry on the bipartite lattices considered here~\cite{Zhang1990,Essler2005}. The latter is generated by the pseudospin operators
\begin{eqnarray}
T^+_j &=& \lr{-1}^{s(j)} \cA{j\downarrow} \cA{j\uparrow}, \label{eq:Tdef}\\
T^-_j &=& \lr{-1}^{s(j)} \cC{j\uparrow} \cC{j\downarrow}, \nonumber\\
T^z_j &=& \lr{n_{j}-1}/2, \nonumber
\end{eqnarray}
which fullfil the SU(2) relations $[T^+_j,T^-_j]=2T^z_j$ and $[T^z_j,T^{\pm}_j]=\pm T^{\pm}_j$. The signs are chosen as $s(j)=0,1$ for sublattice sites A and B, respectively. The global operators $T^x_{\text{tot}}=\sum_j\lr{T^+_j+T^-_j}/2$, $T^y_{\text{tot}}=\sum_j\lr{T^+_j-T^-_j}/(2i)$, $T^z_{\text{tot}}=\sum_jT^z_j$ all commute with the Hamiltonian. Using $\vec{T}_i = (T^x_i,T^y_i,T^z_i)$, the total pseudospin is defined via
\begin{equation}
\avg{\tot{\vec{T}}^2} = \sum_{ij} \avg{\vec{T}_i\cdot\vec{T}_j} = \tot{T}\lr{\tot{T}+1}.
\end{equation}
Note that $\langle T^z_\text{tot}\rangle\sim N_\text{tot}$ up to a constant; half filling corresponds to $\langle T^z_\text{tot}\rangle=0$. The ground state is always in the sector $T_\text{tot}=0$ (we have verified this exemplarily).

\subsection{Heisenberg limit}

For strong onsite interactions $U\to\infty$ the Hubbard model reduces to the Heisenberg chain
\begin{equation}
H_{\text{Heis}} = \sum_{i<j} J_{ij} \vec{S}_i\cdot\vec{S}_j,
\label{eq:Heis}
\end{equation}
where $\vec{S}_i=(S^x_i,S^y_i,S^z_i)$ is the vector of $S=1/2$ spin operators and $J_{ij}=4t_{ij}^2/U$ is the matrix of exchange interactions. This model features a global SU(2) spin symmetry with quantum numbers
\begin{equation}
\sum_{ij} \left\langle\vec{S}_i\cdot\vec{S}_j\right\rangle = \avg{\vec{S}_{\text{tot}}^2} = S_{\text{tot}}\left(S_{\text{tot}}+1\right)
,~M_{\text{tot}} = \sum_i \avg{S^z_i}.
\end{equation}

\subsection{PPP model}

In order to model realistic interactions in polymers, we employ the PPP model~\cite{Ohno1964,Ramasesha1991}
\begin{equation}
H_{\text{PPP}} = H_{\text{Hub}} + \sum_{i<j} V_{ij} \lr{n_i-1}\lr{n_j-1}.
\label{eq:PPP}
\end{equation}
The long-ranged Coulomb interaction parameters are given by the Ohno parametrization
\begin{equation}
V_{ij} = \frac{U}{\kappa\sqrt{1+0.6117R^2_{ij}}},
\label{eq:Vij}
\end{equation}
where $R_{ij}$ is the distance between sites $i$ and $j$ in $\textup{\AA}$ and $\kappa$ regulates the decay of the long-ranged terms.

For the tight-binding hoppings, we use an established parameter set $t_p, t_s, t_d$, see Fig.~\ref{fig:TetrisSketch}. A detailed fitting of the PPP model parameters to optical absorption peaks in PPV yields $t_s=2.2\,\text{eV}$, $t_d=2.6\,\text{eV}$, $t_p=2.4\,\text{eV}$, $U=8\,\text{eV}$, and $\kappa=2$~\cite{Chandross1997A,Chandross1997B}. Carbon atoms in PPV go down a descending armchair path with {120\textdegree} bonds. We follow the authors of Refs.~\cite{Bursill2005,Barford2013} and make an approximation of {180\textdegree} bonds displayed in Fig.~\ref{fig:TetrisSketch}, and we adjust the bond lengths as $R_s=1.283\,\textup{~\AA}$, $R_d=1.194\,\textup{~\AA}$, and $R_p=1.40\,\textup{~\AA}$. Note that the charge-SU(2) symmetry is broken by the PPP model because only the $z$ components of the pseudospin are coupled.

\section{Methods}

\subsection{Noninteracting winding numbers}

We investigate the topological phases of the noninteracting Hamiltonian~\eqref{eq:H0}, which exhibits time-reversal, particle-hole, and sublattice symmetry for all the systems that we study. It can be transformed to a momentum-space representation:
\begin{equation}\label{eq:H0k}
 H_0 = \sum_k H_0(k).
\end{equation}
Due to the sublattice symmetry, $H_0(k)$ can be written in the off-diagonal form
\begin{equation}
 H_0(k) =  
\begin{bmatrix}
0& h(k)^{\dagger}\\
h(k)& 0
\end{bmatrix} .
\end{equation}
The winding number $w$ of $\text{Det} [h(k)]$ around the origin of the complex plane can be used to classify the topological phase~\cite{sch08ClassTI,sch09ClassTI,kitaevTI}. The Hamiltonian~\eqref{eq:H0} is in the CII class of the periodic table of topological insulators with $w\in2\mathbb{Z}$ in one dimension. Since we consider only real hopping parameters, we can omit the spin indices, and in practice consider noninteracting spinless fermions. The corresponding spinless model in the BDI class of topological insulators with $w\in\mathbb{Z}$ in 1D.

\subsection{Tensor networks: DMRG and VUMPS}

In the presence of interactions, we study both finite and infinite molecular chains.
For finite chains, we use the density-matrix renormalization group (DMRG) which is a variational ansatz using matrix-product states (MPS)~\cite{Schollwoeck2011}. For infinite chains, we employ the variational uniform matrix-product state (VUMPS) formalism that works with infinite MPS~\cite{Zauner-Stauber2018}; the numerical unit cell comprises two physical ones (two molecules) unless stated otherwise. MPS-based methods exploit the fact that subsystems are only entangled via their boundary; they are thus highly accurate in one-dimensional gapped systems which have a finite correlation length and are only correlated via a zero-dimensional boundary.

Both the DMRG and the VUMPS algorithm are controlled by the bond dimension $\chi$, a quantity that is related to the number of variational parameters in the MPS ansatz. We carefully ensure that $\chi$ is chosen large enough so that all results shown in this paper are numerically exact (after convergence has been tested, see below).

Our code is able to fully exploit both the spin-SU(2) and charge-SU(2) symmetry of the problem [if present, otherwise U(1) is exploited]. This typically results in an effective gain of $\chi$ by a factor of 2.5-3 per SU(2) symmetry. For the fermionic systems, we shall always work at half filling, which corresponds to $\tot{T}=0$ if the charge-SU(2) symmetry is present, and otherwise to $\tot{N}/L=1\Leftrightarrow\langle\tot{T}^z\rangle=0$. The choice of the spin quantum numbers will be detailed below. Typical values for the effective bond dimensions that are sufficient to obtain numerically-exact results are $\chi_{\text{eff}}\sim 1000$ in Fig.~\ref{fig:YchainPD} or $\chi_{\text{eff}}\sim 3500$ in Fig.~\ref{fig:YchainSzSmol}.

Even though MPS-based methods are the method of choice for the problems considered here, there are two hurdles. Large unit cells, long chains, and long-ranged interactions in combination with local-only updates of the DMRG algorithm may lead to getting caught in local minima. In order to avoid this, we use the two-site algorithm in the initial stages of the DMRG and switch to the one-site algorithm with perturbations~\cite{Hubig2015} in the late stages, and we implement well-converged starting points whenever possible. Moreover, representing the long-ranged PPP as a matrix-product operator (MPO) is challenging; here we use a lossless MPO compression algorithm~\cite{Hubig2017}.

\subsection{Diagnosing the Haldane phase}

The Haldane phase can be numerically diagnosed in various different ways.

\subsubsection{String order}
Sufficient but not necessary is a finite value of the string order parameter~\cite{Pollmann2010}
\begin{equation}
C_{\text{string}}(d) = -\avg{S^z_{i_m}\lr{\prod_{k_m=i_m+1}^{i_m+d-1}e^{i\pi S^z_{k_m}}} S^z_{i_m+d} }
\label{eq:Cstr}
\end{equation}
where the spin operator $S^z_{i_m}$ in a unit cell (i.e., molecule) $i_m$ is defined as
\begin{equation}
S^z_{i_m} = \sum_{i\in\text{unit cell }i_m} S^z_i.
\end{equation}
We compute the string order for infinite systems. The convergence w.r.t.~to the distance $d$ is typically rather quick (see Fig.~\ref{fig:PPVstring} for an example), and we employ a cutoff of $d=d_{\text{max}}=40$. Since Eq.~(\ref{eq:Cstr}) does not have a spin SU(2)-invariant form, we exploit only the spin-U(1) symmetry and focus on the sector $\tot{M}=0$.

\subsubsection{Entanglement spectrum}

A general characteristic of a SPT state is the global even degeneracy of the entanglement spectrum of a bipartition $\ket{\Psi} = \sum_i s_i \ket{\Psi^A_i}\ket{\Psi^B_i}$ at some bond~\cite{Pollmann2010}. We compute the Schmidt values $s_i$, which are granted as a byproduct of any MPS algorithm, for the infinite system and only in the Heisenberg limit. It is known that exploiting the SU(2) spin symmetry can lead to problems~\cite{Li2013,Singh2015}, and we thus employ the spin-U(1) symmetry with $\tot{M}=0$ only. In order to condense the result to a single characterizing number, we plot the staggered sum
\begin{equation}
C_{\text{deg}} = \sum_i \lr{-1}^i s_i,
\label{eq:Cdeg}
\end{equation}
which becomes zero in the case of an even global degeneracy. 

\subsubsection{Edge modes}
For open boundary conditions, the spin gap in the Haldane phase is exponentially small in the system size~\cite{KennedyTasaki1992}, and the singlet ($\tot{S}=0$) and triplet ($\tot{S}=1$) states become degenerate in the thermodynamic limit. A hallmark of the Haldane phase are local magnetic moments $\avg{S^z_{i_m}}$ that are exponentially localized at the edges. We compute $\avg{S^z_{i_m}}$ in the triplet sector with $\tot{M}=1$.

Additional diagnostics involves the Berry phase, which measures the parity of the singlet valence bonds~\cite{Hatsugai2006}, or nonlocal order parameters ~\cite{Pollmann2012,Haegeman2012} that can be directly connected to the protective symmetries. Since both are more complicated to work with, we do not consider them in this paper.

\begin{figure}
\begin{center}
\includegraphics[width=\columnwidth]{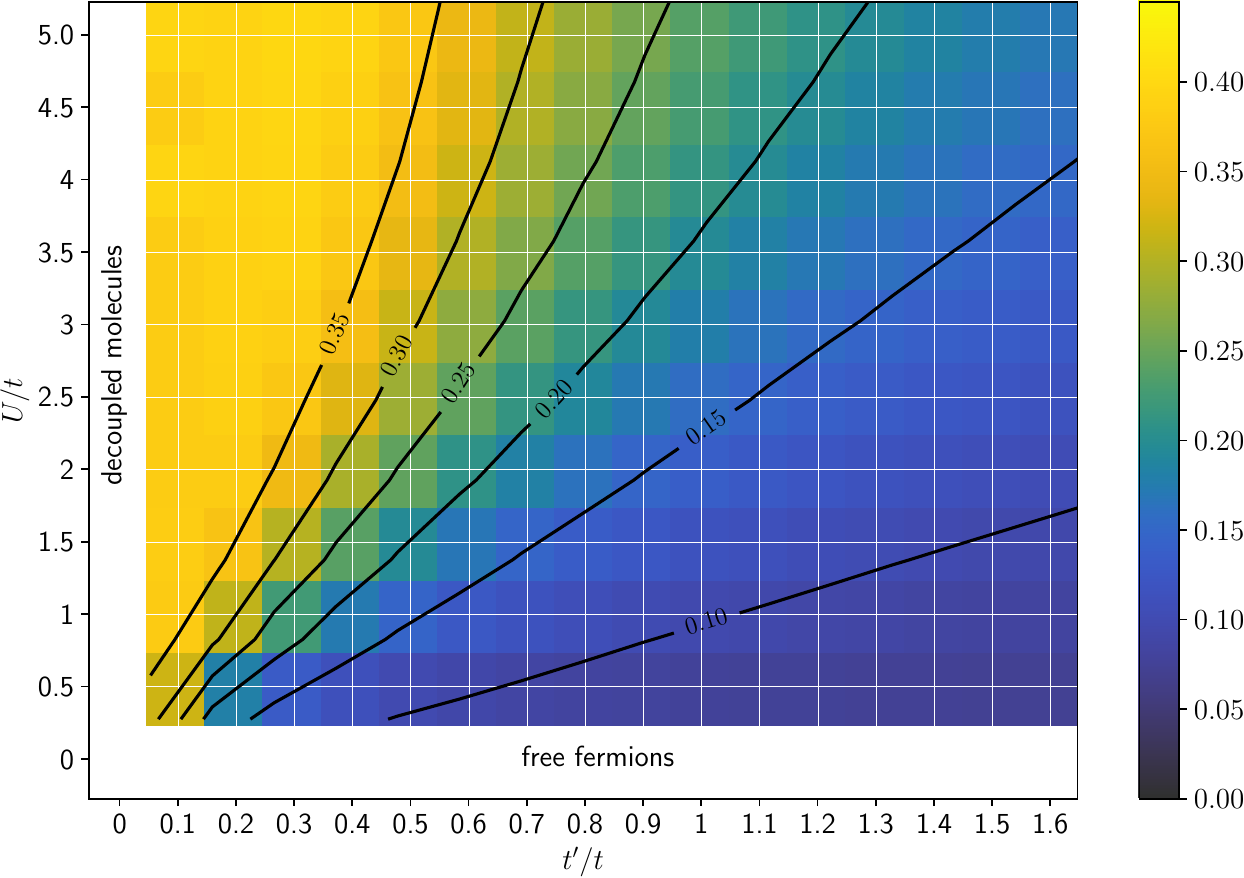}
\caption{
\label{fig:YchainPD}
Phase diagram of the Y-chain. The colormap shows the string order parameter of Eq.~\eqref{eq:Cstr}; it is finite for arbitrary $t'$ and $U$, and the system is thus always in the Haldane phase. The results were obtained via the VUMPS algorithm for an infinite system using spin-U(1)$\times$charge-SU(2) symmetry in the sector $M_{\text{tot}}=0$ at half filling $T_{\text{tot}}=0$.
}
\end{center}
\end{figure}

\section{\label{sec:Ychain}The Y-chain}

We first study the Y-chain, which can be viewed as an approximative coarse-grained model for a chain of triangulenes. Triangulene is a graphene nanoflake with $N_A=12$ ($N_B=10$) carbon atoms in the majority (minority) sublattice (see Fig.~\ref{fig:TetrisSketch}), hence the individual molecule has a ground-state spin of $\tot{S}=1$ within the Hubbard model with $U>0$~\cite{Su2020,Turco2023} by virtue of Lieb's theorem~\cite{Lieb1989}.
For the interaction within triangulene, a rather weak $U/t_1\sim1.0-1.3$ has been suggested that accounts for substrate screening effects~\cite{Mishra2021,Henriques2023}.
When arranged as a chain, the dominant hopping $t_1$ only couples the minority sublattices and produces an unstable flat band. Therefore, it is important to consider a third-neighbor-coupling $t_3/t_1\approx0.1$ (dotted lines in Fig.~\ref{fig:TetrisSketch}), which preserves the sublattice symmetry and lifts the flat-band degeneracy~\cite{Henriques2023}.

Coupled triangulenes are complex to deal with beyond dimers or trimers, but they can be coarse-grained into a Y-shaped molecule with similar properties~\cite{Mishra2021,Catarina2022}. Such a molecule features $N_A=3$, $N_B=1$ and thus also a molecular ground-state spin of $\tot{S}=1$. Fitting the interchain hopping $t'$ of this Y-chain to the full triangulene yields $t'/t\sim 0.17-0.18$~\cite{Mishra2021,Henriques2023}. For both the full system and the Y-chain, an effective low-energy model exists where the molecules are reduced to spin-1 sites. The effective Hamiltonian has a dominating direct exchange term~\cite{Catarina2022,Henriques2023}, and the ground state is in the Haldane phase.

\begin{figure}
\begin{center}
\includegraphics[width=\columnwidth]{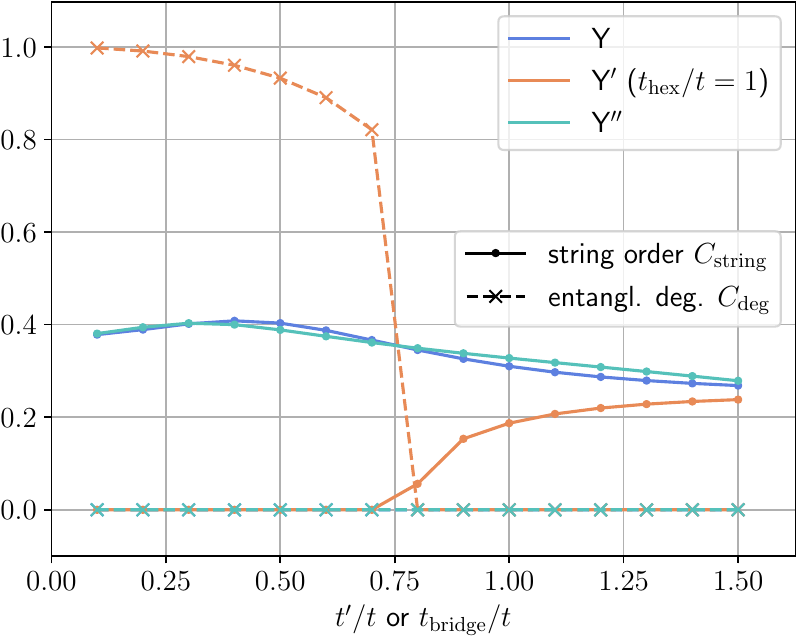}
\caption{
\label{fig:StrongCoupling}
String order parameter $C_{\text{str}}$ defined in Eq.~\eqref{eq:Cstr} and alternating sum of singular values $C_{\text{deg}}$ defined in Eq.~\eqref{eq:Cdeg} for three different tetris chain in the strong-coupling limit $U\to\infty$. The results were obtained within the corresponding Heisenberg model using the VUMPS algorithm; only spin-U(1) symmetry with $M_{\text{tot}}=0$ was exploited. We plot the hopping ratios which are related to the exchange interaction via $t'/t=\sqrt{J'/J}$ (and likewise for $t_{\text{bridge,hex}}$). The Y and Y$^{\prime\prime}$ geometries are always in the Haldane phase, characterized by $C_{\text{str}}>0$ and $C_{\text{deg}}=0$, but there is a phase transition for the Y$^{\prime}$ geometry.
}
\end{center}
\end{figure}

Here, we lift the restriction of small $t'/t$ and $t'/U$ and compute the full $t'-U$ phase diagram within the Hubbard model of Eq.~\eqref{eq:HHub}. Naively, it could be expected that the Haldane phase breaks for large $t'$ since spin-1 molecules cease to be a relevant reference point. This turns out to be incorrect: the string order parameter is finite and the Haldane phase persists in the whole $t'-U$ plane, see Fig.~\ref{fig:YchainPD}.

\begin{figure}[b]
\begin{center}
\includegraphics[width=\columnwidth]{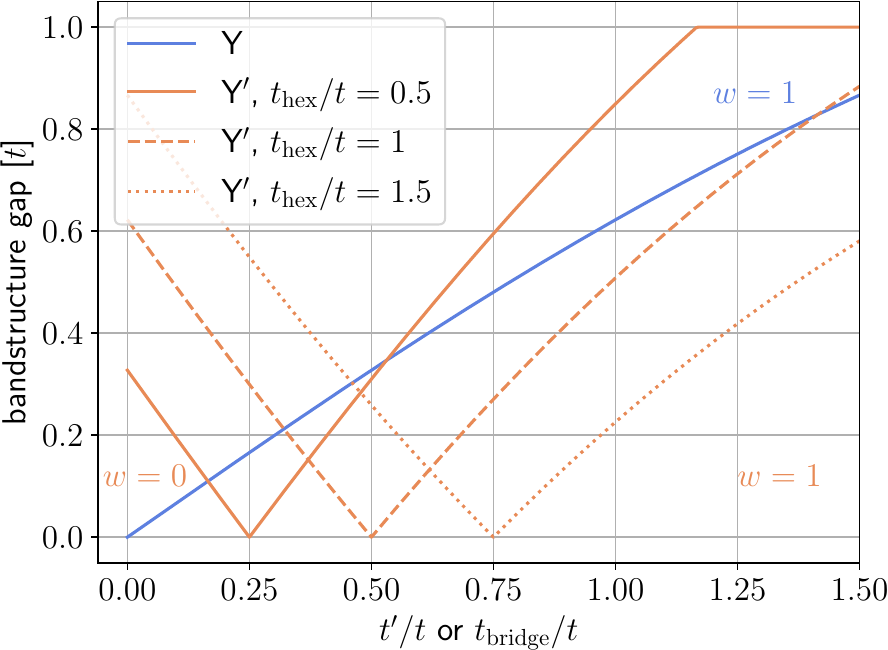}
\caption{
\label{fig:gapBandstructure}
Gap in the single-particle bandstructure and non-interacting winding numbers $w$ at half filling for the Y- and Y$^{\prime}$-chain depicted in Fig.~\ref{fig:TetrisSketch}. A gap closing indicates the transition between the trivial ($w=0$) and topological ($w=1$) phases at $U=0$. The Y-chain is gapped and in a topological phase for any finite $t'$. The Y$^{\prime\prime}$-chain is always gapless.
}
\end{center}
\end{figure}

One can shed more light on this result by looking at the limits of weak and strong interactions. The noninteracting band structure is always gapped and topologically nontrivial with a winding number $w=1$ for any $t'>0$, see Fig.~\ref{fig:gapBandstructure}~\cite{Henriques2023}, which establishes a stable Haldane phase in the weak-$U$ limit for any finite $t'$. In the strong-coupling limit $U\to\infty$, the system reduces to the Heisenberg spin chain of Eq.~\eqref{eq:Heis}, and the corresponding string order parameter as well as the entanglement spectrum degeneracy are shown in Fig.~\ref{fig:StrongCoupling}. The system is always in the Haldane phase irrespective of $t'$. Thus, the Haldane-phase mechanisms 3, 4, and 6 provide cooperating limits on three sides of the phase diagram, stabilizing it everywhere in between.

\begin{figure}
\begin{center}
\includegraphics[width=\columnwidth]{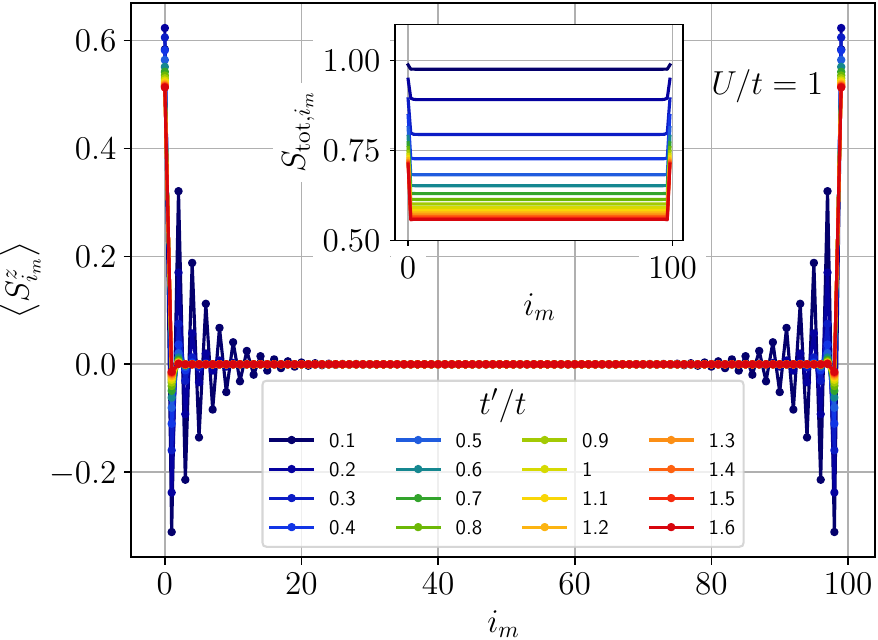}
\caption{
\label{fig:YchainSzSmol}
Magnetization $\avg{S^z_{i_m}}$ of each Y-molecule of a finite chain with $N_{\text{cells}}=100$ (400 sites) for $U/t=1$ and different values of $t'/t$, obtained by a spin-SU(2)$\times$charge-SU(2) DMRG calculation in the triplet sector $\tot{S}=\tot{M}=1$ at half filling $\tot{T}=0$. \textit{Inset:} The total spin per molecule $S_{\text{tot},i_m}$ obtained from Eq.~\eqref{eq:Stotmol}. While local spin-1 moments $S_{\text{tot},i_m}\approx 1$ are only well-defined in the molecular limit $t'/t\ll1$, edge states are localized more strongly for large $t'$ because the gap is increasing.
}
\end{center}
\end{figure}

\begin{figure*}
\begin{center}
\includegraphics[width=\textwidth]{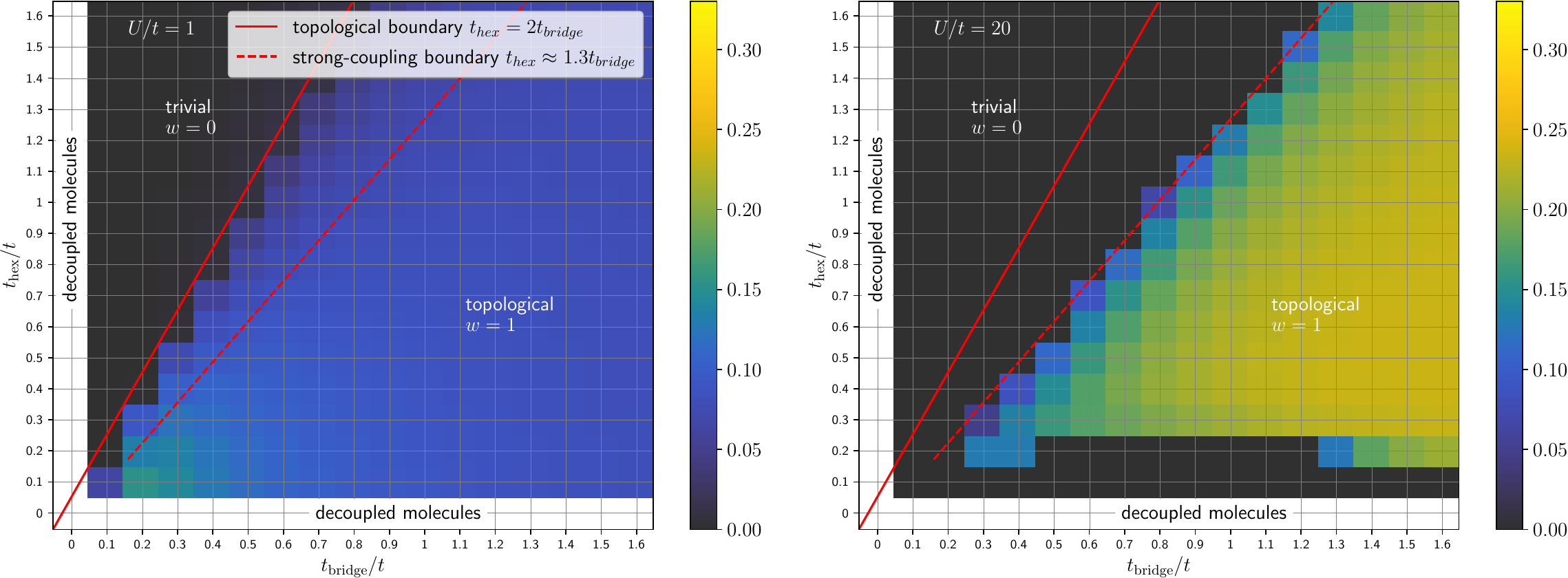}
\caption{
\label{fig:YprimePD}
Phase diagram for the Y$^{\prime}$-chain as a function of $t_{\text{hex}}$ and $t_{\text{bridge}}$ (see Fig.~\ref{fig:TetrisSketch}) for $U/t=1$ (left) and $U/t=20$ (right). The colormap depicts the string order parameter in the infinite system computed using the VUMPS algorithm (symmetry and quantum numbers  are as in Fig.~\ref{fig:YchainPD}). The red solid line shows the small-$U$ result for the boundary between the trivial and the Haldane phase obtained from a bandstructure calculation. The dotted line indicates the large-$U$ phase boundary determined within the corresponding Heisenberg spin model.
}
\end{center}
\end{figure*}

Our results can be further corroborated by looking at finite chains. Fig.~\ref{fig:YchainSzSmol} illustrates that edge states are clearly exponentially localized. The inset shows the total spin per unit cell (i.e., molecule) computed using
\begin{equation}
\sum_{i,j\in\text{unit cell }i_m} \avg{\vec{S}_i\cdot\vec{S}_j} = S_{\text{tot},i_m}\lr{S_{\text{tot},i_m}+1}.
\label{eq:Stotmol}
\end{equation}
Interestingly, one can see that local spin-1 moments $S_{\text{tot},i_m}\approx 1$ are only well-defined in the molecular limit $t'/t\ll1$, illustrating that such spin-1 moments are not a necessary feature of a Haldane phase. In contrast, edge states are actually localized more strongly for large $t'$ because the gap is increasing.

The Haldane phase manifests for arbitrary values of $t'$, and triangulene chains are hence not the only pathway to obtain this phase. E.g., the homogeneous case $t'=t$ of the Y-chain describes a depleted Hubbard ladder (see Fig.~\ref{fig:TetrisSketch}), which could be implemented using designer lattices~\cite{Huda2020}.

\section{\label{sec:Yprime}The Y$^{\prime}$-chain}

We now investigate another way to couple the Y-molecules, the ``Y$^{\prime}$-chain''. It features two different intermolecular couplings $t_{\text{hex}}$ and $t_{\text{bridge}}$, see Fig.~\ref{fig:TetrisSketch}. The interaction is described using the Hubbard model of Eq.~\eqref{eq:HHub}. The unit cell includes $L_{\text{cell}}=8$ sites. It is instructive to first study simple limiting cases before looking at the full phase diagram.

Only if both $t_{\text{hex}} \ll t$ and $t_{\text{bridge}} \ll t$ do we have well-defined spin-1 molecules via Lieb's theorem. The resulting model should have dimerized couplings, so that it is not immediately clear whether it is in the Haldane phase. Deriving the effective spin model is beyond the scope of this paper. If either $t_{\text{hex}} \ll t$ or $t_{\text{bridge}} \ll t$, we have weakly coupled 8-site molecules with balanced sublattices $N_A=N_B$ and no effective spin-1 model.

The limit of small $U$ can again be addressed by looking at the noninteracting winding number. We find a trivial value of $w=0$ for $t_{\text{hex}}>2t_{\text{bridge}}$ and a nontrivial $w=1$ for $t_{\text{hex}}<2t_{\text{bridge}}$ separated by a gap closure (see Fig.~\ref{fig:gapBandstructure}). This shows that the Haldane manifests at small $U$ for $t_{\text{hex}}<2t_{\text{bridge}}$.

In the strong-coupling limit of large $U$, the Y$^\prime$-chain can be described by a Heisenberg model on the same geometry with $J=4t^2/U$, $J_{\text{hex}}=4t_{\text{hex}}^2/U$, and $J_{\text{bridge}}=4t_{\text{bridge}}^2/U$. Fig.~\ref{fig:StrongCoupling} shows the corresponding string order parameter as well as the entanglement degeneracy for $t_\text{hex}/t=1$. The Y$^\prime$-chain is only in the Haldane phase beyond a critical value of $1.3t_{\text{bridge}}\gtrsim t_{\text{hex}}$, which is different from the small-$U$ result $2t_{\text{bridge}}>t_{\text{hex}}$. We hypothesize that this remains true for any value of $t_\text{hex}/t$, which we will confirm explicitly for the full model at large $U$.

After studying these simple limits, we now turn to the full phase diagram, which is shown in Fig.~\ref{fig:YprimePD} in the $t_{\text{bridge}}-t_{\text{hex}}$ plane for two different values of $U$. For weak coupling $U/t=1$, the transition into the Haldane phase occurs around $2t_{\text{bridge}}\approx t_{\text{hex}}$ in agreement with the noninteracting bandstructure calculation (mechanism 4). In the strong-coupling case $U/t=20$, the results are consistent with the Heisenberg-model prediction $1.3t_{\text{bridge}}\approx t_{\text{hex}}$ (mechanism 6). In between $t_{\text{hex}}=2t_{\text{bridge}}$ and $t_{\text{hex}}\approx1.3t_{\text{bridge}}$, the mechanisms 4 and 6 compete with each other. We observe that the same holds true for small $t_{\text{hex}}/t$ and $t_{\text{bridge}}/t\sim 0.9\pm0.45$ (see the center bottom of the phase diagram for $U/t=20$). In both of these regions, the existence of the Haldane phase depends on $U$, and it recedes as $U$ is increased despite the nontrivial topological bandstructure.

We observe that a finite region around the homogeneous point $t_{\text{hex}}\approx t_{\text{bridge}}$ is always in the Haldane phase for any value of $U$. One can thus suspect that the near-homogeneous electronic model of PPV should be in the Haldane phase as well. We will investigate this in more detail in Sec.~\ref{sec:PPV}.

\section{\label{sec:Yprimeprime}The Y$^{\prime\prime}$-chain}

We briefly comment on the Y$^{\prime\prime}$-chain, see Fig.~\ref{fig:TetrisSketch}. We find that its noninteracting bandstructure is always gapless, and one cannot gain any further insights from this limit. For strong coupling, the system is in the Haldane phase for any $t'$ (see Fig.~\ref{fig:StrongCoupling}). We speculate that the Y$^{\prime\prime}$-chain becomes a Mott insulator for any value of $t'$, but we refrain from investigating this in more detail since it seems unlikely that it can be engineered as a real system.

The Y$^{\prime\prime}$ coupling can be viewed as a molecular limit of the Hubbard ladder; it hence provides an alternate route of transforming a Hubbard ladder to a spin model via weakening the $t'$ bonds. This path entirely avoids the strong-coupling limit of large $U$ which is normally considered in studying the ladder~\cite{Anfuso2007,Moudgalya2015}. This also illustrates that in order to detect features of the Haldane phase in a finite Hubbard ladder, it is necessary that two end atom are attached at the ends (see Fig.~\ref{fig:StrongCoupling}).

We note that similar arguments can be found in Ref.~\cite{White1996} for a $S=1/2$ Heisenberg ladder, which can be transformed smoothly into a diagonally-coupled ladder that features a Haldane phase; additional edge sites naturally appear also in this system.

\begin{figure}
\begin{center}
\includegraphics[width=\columnwidth]{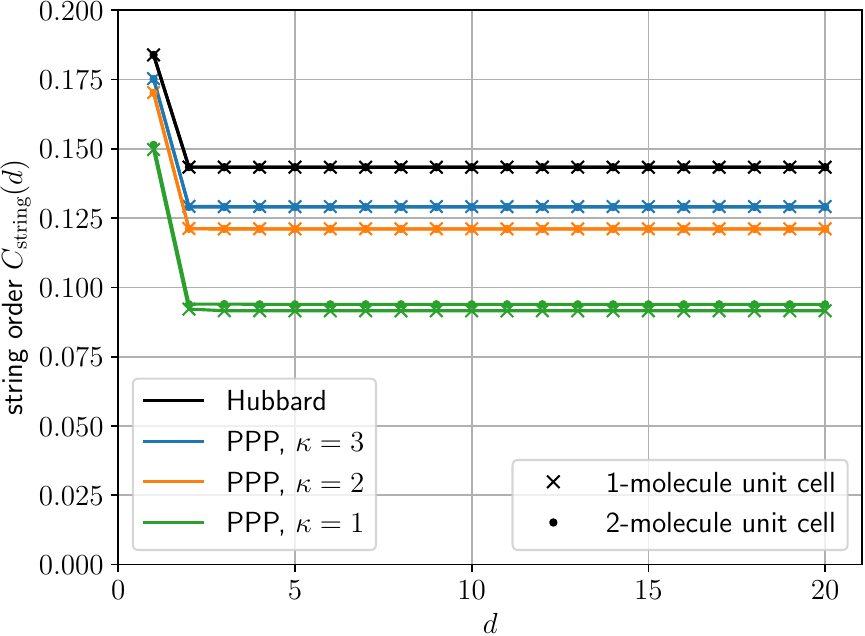}
\caption{
\label{fig:PPVstring}
String order parameter of Eq.~\eqref{eq:Cstr} as a function of $d$ for an infinite PPV chain with $t_s=2.2\,\text{eV}$, $t_d=2.6\,\text{eV}$, $t_p=2.4\,\text{eV}$, and $U=8\,\text{eV}$ (see Fig.~\ref{fig:TetrisSketch}) at half filling and $\tot{M}=0$. The results were obtained using the VUMPS algorithm; spin-U(1)$\times$charge-SU(2) were exploited for the Hubbard model, and spin-U(1) $\times$ charge-U(1) symmetries were exploited for the PPP model. The numerical unit cells comprises one or two physical unit cells, which limits the range of the PPP interactions to $d_V=7$ and $d_V=15$ sites, respectively.
}
\end{center}
\end{figure}

\section{\label{sec:PPV}PPV}

The PPV polymer has received a lot of attention since the discovery of electroluminescence in 1990~\cite{Burroughes1990,Soos1993,Chandross1997A,Chandross1997B,Chandross1997C,Chandross1997D,Bursill1998,Bursill2002,Muellen2023} and is used, e.g., to engineer light-emitting diodes~\cite{Shinar2004BOOK}. To the best of our knowledge, the possibility of using this system to realize the Haldane phase has not been investigated so far.

We first compute the winding number of the noninteracting bandstructure as a function of the couplings $t_p$, $t_s$, and $t_d$ (see Fig.~\ref{fig:TetrisSketch}). The trivial and topological phases manifest for $2t_d t_p>t_s^2$ and $2t_d t_p<t_s^2$, respectively. The PPV polymer is governed by near-homogeneous values $t_d\approx t_s\approx t_d$, and we thus expect it to be in the Haldane phase in the weak-coupling limit of small interactions by virtue of the mechanism 4.

We now investigate whether or not the Haldane phase persists in the presence of more realistic long-range PPP interactions of Eq.~\eqref{eq:PPP} whose range we limit to $d_V$ sites in order to perform numerics. We first compute the string order parameter using the VUMPS algorithm and employ numerical unit cells of 8 and 16 sites which entails $d_V=7$ and $d_V=15$ sites, respectively. The result is shown in Fig.~\ref{fig:PPVstring} for different $\kappa$, and Hubbard-model data is included for reasons of completeness. The PPP interactions become more long-ranged with decreasing $\kappa$, but the system remains robustly in the Haldane phase. This is supported by the entanglement spectrum (data not shown).

\begin{figure}
\begin{center}
\includegraphics[width=\columnwidth]{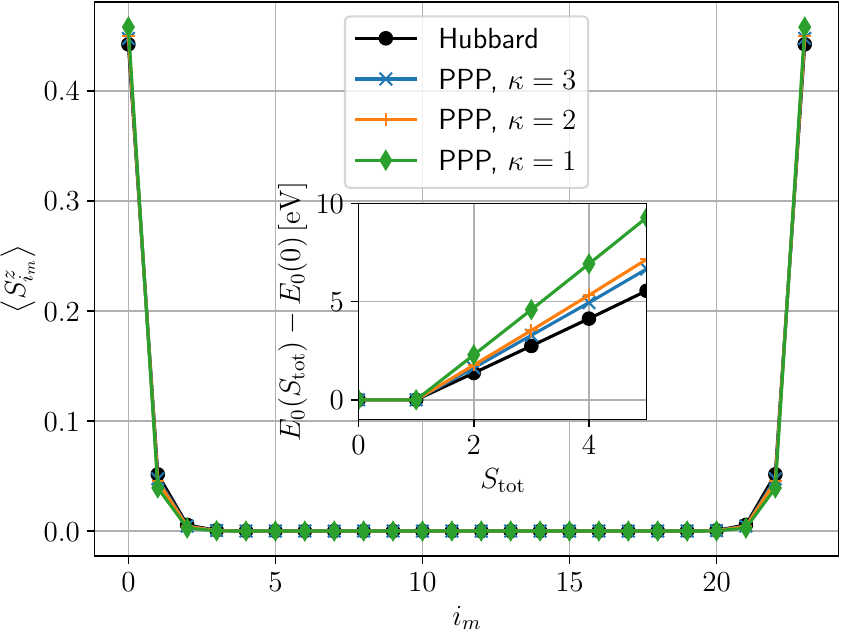}
\caption{
\label{fig:PPVfinite}
Magnetization $\avg{S^z_{i_m}}$ in the triplet sector $\tot{S}=\tot{M}=1$ at half filling as a function of the unit cell (molecule) $i_m$ of a finite PPV chain with $N_{\text{cells}}=24$ (192 sites). The data was obtained from spin-SU(2)$\times$charge-SU(2) and spin-SU(2)$\times$charge-U(1) DMRG calculations for the Hubbard and the PPP model, respectively. \textit{Inset:} The lowest energy $E_0\lr{\tot{S}}$ in a given sector of the total spin $\tot{S}$ shifted by $E_0\lr{\tot{S}=0}$.
}
\end{center}
\end{figure}

\begin{figure}
\begin{center}
\includegraphics[width=\columnwidth]{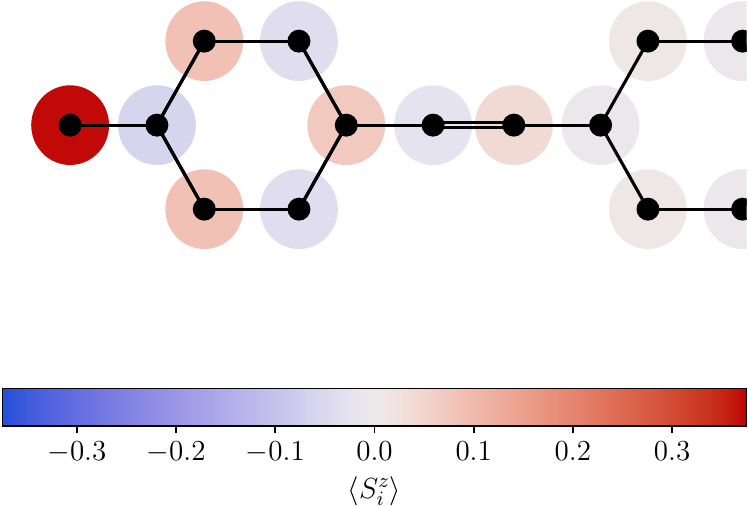}
\caption{
\label{fig:PPV_SzGeometry}
Real-space texture of the magnetization $\avg{S^z_{i}}$ in the triplet sector $\tot{S}=\tot{M}=1$ for the PPP chain with $\kappa=2$.
}
\end{center}
\end{figure}

We corroborate these results for finite systems and $d_V=15$. Figure~\ref{fig:PPVfinite} shows the local spin at each unit cell $\avg{S^z_{i_m}}$ for $N_{\text{cells}}=24$ unit cells. The exponentially-localized edge state can be identified clearly, and it is robust with respect to the PPP interactions.  Moreover, the singlet-triplet gap vanishes (inset). Finally, Fig.~\ref{fig:PPV_SzGeometry} shows that within the unit cell, most of the excess spin is localized on the outermost carbon atom. In sum, the finite-system data is clearly consistent with a Haldane phase.

We note that in order to observe the Haldane edge states, two carbon atoms need to be added to both ends of the chain, see Fig.~\ref{fig:TetrisSketch}. This is in contrast to the more common boundary condition that terminates the chain at the vinyl rings.

\section{\label{sec:Conclusion}Conclusion}

The Y-chain constitutes a simplified model of coupled Y molecules that each have a total spin of $\tot{S}=\big|N_A-N_B\big|/2=1$ by Lieb's theorem and that hence resemble triangulene. In this paper, we have extended the previous studies of the Y-chain and computed the full phase diagram. We find that the Haldane phase fills out the whole $t'-U$ plane including the homogeneous limit $t'=t$. This offers an alternative route to realize this phase, e.g., using designer lattices~\cite{Huda2020}. Such a route is in fact advantageous in the sense that the noninteracting protective gap is on the scale of the main hopping $t\sim2-3\,\text{eV}$ (see Fig.~\ref{fig:gapBandstructure}) and not on the scale of the intermolecular hopping $t'$ or the exchange interaction $t'^2/U$, which are both small in realistic scenarios. Correspondingly, the magnetic edge states are strongly localized (see Figs.~\ref{fig:YchainSzSmol} and \ref{fig:PPV_SzGeometry}).

The Y-molecules can be coupled in various alternative ways. Two important restrictions are that the sublattice label must shift from one molecule to the next so that a global balance $N^{\text{tot}}_A=N^{\text{tot}}_B$ is maintained (otherwise a global ferromagnetic state would be induced~\cite{Essalah2021}) and that the majority sublattices of each molecule have to be coupled in order to prevent a flat band.

This leads to the concept of \textit{tetris chains} (see Fig.~\ref{fig:TetrisSketch}): The Y$^{\prime}$-chain exhibits a more complex phase diagram, and the Haldane phase exists only in certain regions. For near-homogeneous couplings, one obtains a chain of PPV polymers, and despite the multitude of studies devoted to this system, it has to the best of our knowledge not been realized that it features a Haldane phase which remains stable even in the presence of long-ranged PPP interactions.

Another tetris chain is the Y$^{\prime\prime}$ one, which is related to the site-edged Hubbard ladder and which thus provides an alternate route of transforming the Hubbard ladder to a spin model via the mechanism 3 without having to resort to the strong-coupling limit~\cite{Anfuso2007,Moudgalya2015}.

A coupling geometry that we have not considered in this paper is a stacked arrangement of the Y-molecules. This is relevant experimentally since graphene nanoflakes can also be stacked~\cite{Guo2022}. We find that this coupling leads to a gapless noninteracting bandstructure and thus expect a Mott insulator for $U>0$. It remains an open question whether such a system would qualitatively behave as the Hubbard ladder or if some interesting cluster Mott insulators~\cite{Hickey2016,Nourse2021,Jayakumar2023} can be induced. This is left for future work.

\subsection*{Acknowledgements}
Discussions with Eric Jeckelmann are gratefully acknowledged.


\bibliography{HaldaneTetrisBib.bib}

\end{document}